\documentclass[runningheads]{llncs}
\usepackage[T1]{fontenc}
\usepackage{graphicx}
\usepackage{caption}
\usepackage{subcaption}
\usepackage{hyperref}
\usepackage{amssymb}
\usepackage{amsmath}
\usepackage[utf8]{inputenc}
\usepackage{gensymb}  
\begin{document}
\title{Adversarial Robustness of MR Image Reconstruction under Realistic Perturbations}
%
%
\author{
Jan Nikolas Morshuis \inst{1} \and Sergios Gatidis \inst{2} \and
Matthias Hein \inst{1} \and
Christian F. Baumgartner \inst{1}}

\authorrunning{JN Morshuis et al.}
\titlerunning{Adversarial Robustness of MRI Recon. under Realistic Perturbations}
%
\institute{
Cluster of Excellence Machine Learning, University of Tübingen \and Max-Planck Institute for Intelligent Systems \\
\email{nikolas.morshuis@uni-tuebingen.de}}

\maketitle             

\begin{abstract}
Deep Learning (DL) methods have shown promising results for solving ill-posed inverse problems such as MR image reconstruction from undersampled $k$-space data. However, these approaches currently have no guarantees for reconstruction quality and the reliability of such algorithms is only poorly understood. Adversarial attacks offer a valuable tool to understand possible failure modes and worst case performance of DL-based reconstruction algorithms. In this paper we describe adversarial attacks on multi-coil $k$-space measurements and evaluate them on the recently proposed E2E-VarNet and a simpler UNet-based model. In contrast to prior work, the attacks are targeted to specifically alter diagnostically relevant regions. Using two realistic attack models (adversarial $k$-space noise and adversarial rotations) we are able to show that current state-of-the-art DL-based reconstruction algorithms are indeed sensitive to such perturbations to a degree where relevant diagnostic information may be lost. Surprisingly, in our experiments the UNet and the more sophisticated E2E-VarNet were similarly sensitive to such attacks. Our findings add further to the evidence that caution must be exercised as DL-based methods move closer to clinical practice. 

\end{abstract}
\section{Introduction}

MR image acquisition is notoriously limited by long scan times adding to patient discomfort and healthcare costs. By acquiring only a subset of measurements in the $k$-space this process can be substantially accelerated. However, reconstructing images from such undersampled $k$-space data requires solving an ill-posed inverse problem. The problem was initially approached using techniques rooted in the theory of compressed sensing which offers provable error bounds (see e.g.~\cite{rudin1992nonlinear}). In recent years, it has been shown that MRI reconstruction algorithms based on deep learning (DL) can substantially outperform classical reconstruction methods both in accuracy and speed~\cite{results_fastmri_2020,zhu2018image,schlemper2017deep,end-to-end-varnet,unet_reconstruction,tezcan2018mr}. Such approaches, however, currently lack provable robustness guarantees. Moreover, their failure cases are not sufficiently understood, putting into question the trustworthiness of such approaches in real-world scenarios.

Recently, there has been an increasing focus on better understanding the robustness properties of DL-based reconstruction algorithms. For instance, in the 2020 fastMRI challenge it was reported that top performing deep learning based methods can sometimes miss small abnormalities at $8\times$ acceleration~\cite{results_fastmri_2020,caliva2020adversarial}. A priori it is unclear whether the information of these abnormalities is lost in the $k$-space subsampling step, which would make a correct reconstruction physically impossible, or whether the information is contained in the measurements but reconstruction algorithms fail to retrieve it. Cheng et al.~\cite{cheng2020} provided compelling evidence that in many cases the latter is the case by inserting adversarially generated false positives which fail to be reconstructed. This suggests that robustness properties of DL-based reconstruction algorithms deserve closer scrutiny. In another line of work, a large analysis of the 2019 fastMRI challenge submissions~\cite{robustness_fastmri_2019} found that the performance of DL-based reconstruction methods can suffer if a lower SNR is encountered during test-time than during training time. Darestani et al.~\cite{darestani2021measuring} further showed a sensitivity to different acquisition techniques and different anatomies.

While it seems well documented that clinically relevant failure modes do exist, comparatively little work has been dedicated to understanding the worst case performance. Better understanding the maximum harm an algorithm can do is crucial for any clinical application and an integral part of the medical auditing process~\cite{liu2022medical}. In absence of theoretical guarantees a number of works have attempted to provide empirical bounds on the worst case performance using adversarial attacks. While it is well established that DL-based classification networks are very sensitive to such attacks due to the ``vulnerable'' decision boundary stemming from the discrete nature of the problem, MR reconstruction algorithms do not necessarily suffer from the same problem~\cite{Antun,solving_inverse_problems}. Firstly, MR reconstruction can be seen as a regression problem lacking discrete boundaries. Furthermore, many DL-based algorithms, such as the state-of-the-art E2E-VarNet (E2E-VN)~\cite{end-to-end-varnet}, which we study in this paper, are inspired by classical algorithms with documented robustness properties.

Antun et al.~\cite{Antun} demonstrated that, indeed, small adversarial perturbations applied to the input of reconstruction algorithms may result in severe reconstruction artifacts. Similarly, Darestani et al.~\cite{darestani2021measuring} investigated adding adversarial noise to the $k$-space measurements, observing that DL-based methods can be sensitive to the adversarial noise and that such attacks can also substantially degrade diagnostically relevant regions. Genzel et al.~\cite{solving_inverse_problems} also investigated adversarial attacks with bounded noise on the $k$-space measurements. Interestingly, the authors came to the conclusion that the investigated reconstruction networks are relatively stable to such attacks, though results varied with different undersampling patterns. In particular, the authors note that diagnostically relevant regions remain largely unaffected. We thus observe that the evidence for adversarial robustness of MR reconstruction algorithms remains inconclusive. 

In this paper, we aim to shed further light on this question by analysing adversarial robustness on the commonly investigated fastMRI knee dataset~\cite{fastmri,fastmri_plus}. In contrast to prior work we show that when attacking state-of-the-art MRI reconstruction algorithms in a way targeting regions with diagnosed pathologies directly, very small levels of adversarial noise on the $k$-space (of comparable magnitude to the thermal noise always present in MR acquisition) can substantially degrade those regions and in some instances alter diagnostically relevant features. Going beyond the standard noise based adversarial attacks, we furthermore show that, somewhat worryingly, very small adversarial rotations can have similar effects. These findings add additional evidence that the worst case performance of DL-based MR reconstruction algorithms may currently be unacceptable for clinical use. 

    

\section{Methods}

\subsubsection{Problem Setting}

In MR image acquisition typically multiple receiver coils are used to measure the data necessary to produce an image. Each coil $i$ produces a $k$-space measurement $\boldsymbol{k}_i \in \mathbb{C}^n$. Given $\boldsymbol{k}_i$, a coil image $\boldsymbol{x}_i \in \mathbb{C}^n$ can be calculated using the  inverse Fourier transform, such that 
\begin{equation}
    \boldsymbol{x}_i = \mathcal{F}^{-1}(\boldsymbol{k}_i+\boldsymbol{z}),
\end{equation}
where $\boldsymbol{z} \in \mathcal{N}(0,\sigma\cdot \boldsymbol{I})$ is additive Gaussian thermal noise which is unavoidable in MR image acquisition~\cite{noise_in_mri}. In the fastMRI setting only a subset of the $k$-space data is measured. This can be simulated by applying a mask function $\mathcal{M}$ that sets the unmeasured part of the $k$-space data to 0, i.e. $\mathcal{M}(\boldsymbol{k}) = \boldsymbol{k} \odot M$, for a mask $M$ and with $\odot$ the element-wise matrix multiplication. We employ the usual Cartesian undersampling pattern also used for the fastMRI challenge in all our experiments. Thus, a single undersampled coil-image can be expressed as
\begin{equation}
    \tilde{\boldsymbol{x}_i} = \mathcal{F}^{-1}(\mathcal{M}(\boldsymbol{k}_i+\boldsymbol{z})). 
    \label{eq:kspace_to_coilimage}
\end{equation}

The acquired coil-images $\boldsymbol{x}_i$ or $\boldsymbol{\tilde{x_i}}$ can be combined using the voxelwise root-sum-of-squares (RSS) method \cite{rss} to obtain the fully-sampled image $\boldsymbol{X}$ (or the undersampled image $\boldsymbol{\tilde{X}}$, respectively):
\begin{equation}
    \boldsymbol{X} = \sqrt{\sum_{i=0}^{N} |\boldsymbol{x}_i|^2},
    \label{eq:image_calc}
\end{equation}
where $N$ indicates the number of coils.

MR reconstruction algorithms aim to reconstruct $\boldsymbol{X}$ given only the undersampled $k$-space data $\mathcal{M}(\boldsymbol{k}_i$) for $i \in \{1,2, ..., N\}$.  DL-based approaches either start by obtaining the image $\boldsymbol{\tilde{X}}$ created from undersampled measurements according to Eq. \eqref{eq:image_calc} and treat the reconstruction as a de-aliasing problem (e.g.~\cite{unet_reconstruction,schlemper2017deep}, or use the $k$-space data as input directly (e.g. \cite{end-to-end-varnet,zhu2018image}). Here, we define the reconstruction network $f$ as a mapping from (partial) $k$-space to image space. 

\subsubsection{Investigated Reconstruction Methods}\label{sec:recon_methods}

In this paper we investigated two different neural network architectures. Namely a method based on the image-to-image UNet approach proposed in~\cite{unet_orig} as well as the iterative network E2E-VN~\cite{end-to-end-varnet}, which reported state-of-the-art performance on the fastMRI challenge data. Although pretrained weights for both these models exist in the fastMRI-repository \cite{results_fastmri_2020}, they were trained using both train and validation sets of the fastMRI data, and evaluated on a private challenge test set. Since the ground-truth data of the private test set is not publicly available, we conduct our analysis of the model robustness on the validation set. The models therefore had to be retrained using data from the training set only.

Both networks have been trained using a batch-size of 32, like in their original implementation \cite{results_fastmri_2020}. To train with the original batch-size, the E2E-VN requires 32 GPUs of 32 GB memory each, which was not feasible with our local compute infrastructure. This necessitated adapting the network size such that it could fit into the 11GB memory of our local GPUs. For the E2E-VN, we reduced the number of unrolled iterations from 12 to 8 and halved the number of channels in each UNet. The size of the UNet for estimating the sensitivity maps was also reduced from 8 top-level channels to 6. Due to the computational constraints, we also halved the number of channels in the baseline UNet approach. The final training of our E2E-VN adjustment was carried out with 32 GPUs in parallel and took around one day to train, our adjusted UNet used 8 GPUs in parallel and also took around one day to train. Other than the reduction of the number of parameters, our version of the two networks is identical to the originals used for the fastMRI challenge and we expect our insights into its robustness properties to carry over to their larger counterparts. 


\subsubsection{Adversarial Attacks on the Reconstruction Methods}
\label{sec:adversarial_attacks}

When using adversarial attacks, we are interested in finding a worst-case perturbation according to some attack model and within given boundaries. In this paper we explore two types of attack: 1) Adversarial noise added to the $k$-space measurements, which is bounded by a maximum $L_2$-norm, 2) Adversarial rotations where we simulate a slightly rotated position of the patient during acquisition. In the following both methods are described in more detail.

For the first attack model, we aim to find bounded adversarial noise-vectors $\boldsymbol{z}_i$ for every $k$-space measurement $\boldsymbol{k}_i$ of every coil $i$, such that the resulting output changes as much as possible in a user-defined target region. More specifically, we have the following optimization problem:
\begin{equation}
\label{eq:noise_opt}
    \max_{z} ||S\odot\left[f(\mathcal{M}( \boldsymbol{k}+\boldsymbol{z}))-\boldsymbol{X}\right]||_2 \ : \ ||\boldsymbol{z}_i||_2 \le \eta \ ||\boldsymbol{k}_i||_2.
\end{equation}
Here, $\boldsymbol{k}$ is the vector of all $k$-space data $\boldsymbol{k}_i$ from all coils $i$, $\eta$ defines the $L_2$-adversarial noise level relative to every coil measurement $\boldsymbol{k}_i$, and
$S$ is a binary region selection mask to limit the attack to diagnostically relevant regions. Specifically, we use the bounding boxes indicating pathologies provided by the fastMRI+ dataset~\cite{fastmri_plus} to define $S$. In contrast to \cite{solving_inverse_problems,darestani2021measuring}, we restrict the noise-vector $\boldsymbol{z}_i$ for every coil individually instead of the complete measurement, thereby allowing for differences in the coil-sensitivities. 

The proposed realistic rotation attacks can be expressed similarly as
\begin{equation}
\label{eq:adv_rotation}
    \max_{\theta} ||S\odot\left[(\mathcal{R}_{\theta}^{-1}\circ f \circ \mathcal{M}\circ\mathcal{R}_{\theta})(\boldsymbol{k})-\boldsymbol{X}\right]||_2 \ : \ d \in [-\theta_{max}, \theta_{max}].  
\end{equation}
Here, we introduce an operator $\mathcal{R}_{\theta}$ which denotes a rotation of the MRI data by $\theta$. In practice, we rotate the data in the spatial domain and then Fourier-transform back to obtain $\boldsymbol{k}$ similar to \cite{mraugment}. After a prediction is created by the neural network, the inverse rotation $\mathcal{R}_{\theta}^{-1}$ is used to transform the image back. It is then compared to the original untransformed target-image.

The optimization process in Eq. \eqref{eq:noise_opt} was implemented using 10-step projected gradient descent. In every step, the adversarial method adds a noise-vector $\boldsymbol{z}_i$ to the $k$-space $\boldsymbol{k}_i$ for every coil $i$, using this transformed $k$-space data as a new input for the reconstruction method $f$. After the calculation of the main objective in Eq. \eqref{eq:noise_opt} and successive backpropagation, the resulting gradients are added to the noise-vector using a step-size of 0.5. The noise-vector $\boldsymbol{z_i}$ is then renormalized to stay within the pre-defined boundaries. The worst-case rotation angle $\theta$ in Eq.~\eqref{eq:adv_rotation} was found using a grid search using an evenly spaced grid from $-\theta_{max}$ to $\theta_{max}$ with a step-size of $0.1\degree$. Our code is publicly be available online\footnote[1]{\url{https://github.com/NikolasMorshuis/AdvRec}}.

\section{Experiments and Results}

We investigated adversarial noise and rotation attacks as described in Section \ref{sec:adversarial_attacks} on the fastMRI knee dataset~\cite{fastmri}. We explored attacking the entire center-cropped image, as well as targeted attacks on diagnostically relevant regions, which we obtained from the pathology annotations in the fastMRI+ dataset~\cite{fastmri_plus}. We evaluated the adversarial robustness for the E2E-VN~\cite{end-to-end-varnet} as well as the UNet~\cite{unet_reconstruction} and for acceleration factors of $4\times$ and $8\times$.

To explore attacks of varying severity, we repeated all experiments for a range of parameters $\eta$ and $\theta_{max}$. Similar to \cite{solving_inverse_problems} and \cite{darestani2021measuring}, we set the upper bound of the noise vector $\eta$ relative to the $k$-space norm $||\boldsymbol{k}_i||_2$. We explored a range of $\eta \in [0\%, 2.5\%]$. We empirically verified that the adversarial noise levels have a similar scale to thermal noise in our data, by measuring the real noise in the background regions of the input images as described in \cite{firbank1999comparison}. This makes it plausible that similar noise could also occur in practice. For the maximum rotation parameter $\theta_{max}$ we explored a range of $[0\degree, 5\degree]$. 

\begin{figure}[ht!]
     \centering
     \begin{subfigure}[b]{0.48\textwidth}
         \centering
         \includegraphics[width=\textwidth]{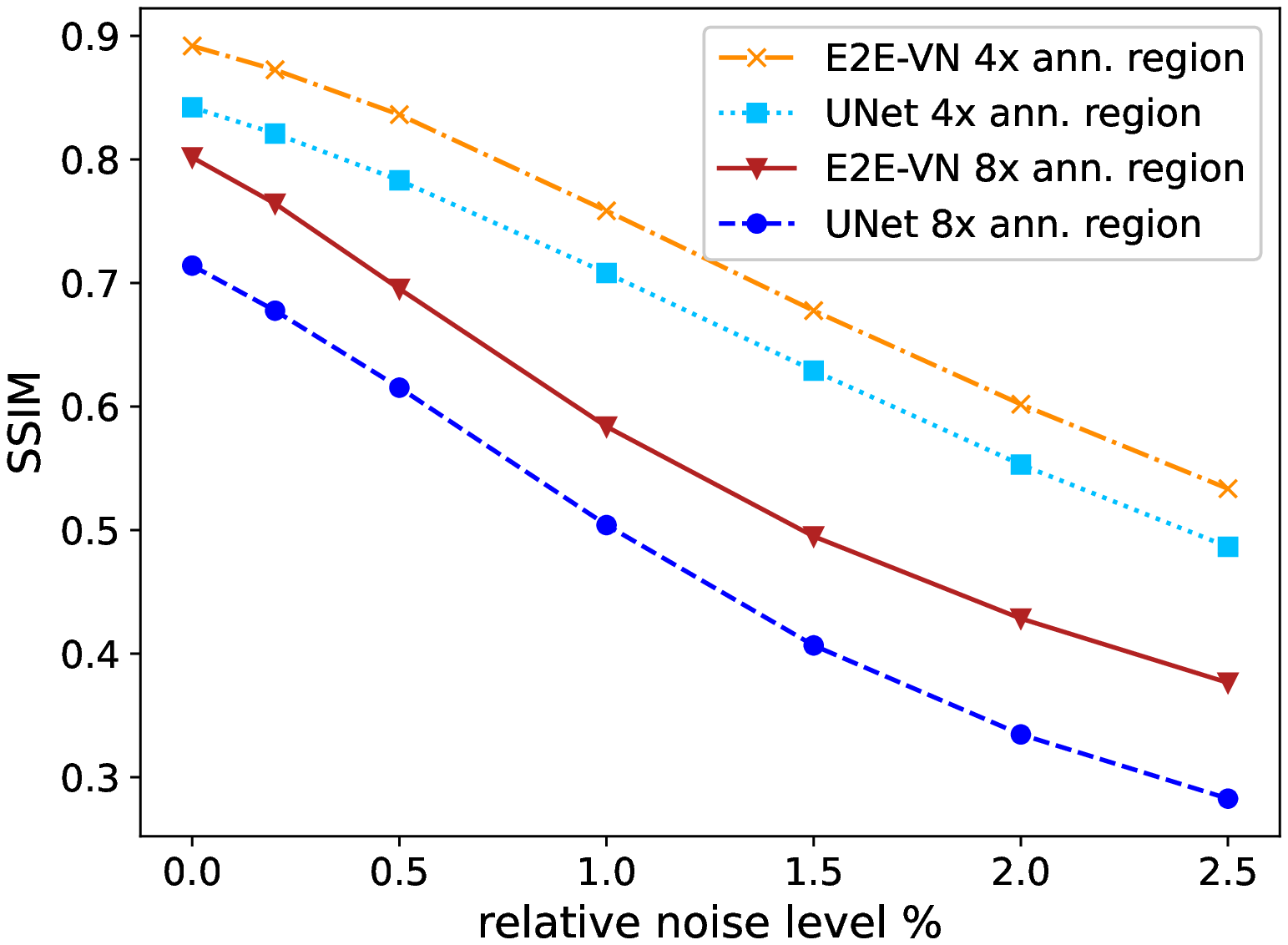}
         \caption{Comparison of acceleration factors}
         \label{fig:SSIM-rel-error}
     \end{subfigure}
     \begin{subfigure}[b]{0.48\textwidth}
         \centering
         \includegraphics[width=\textwidth]{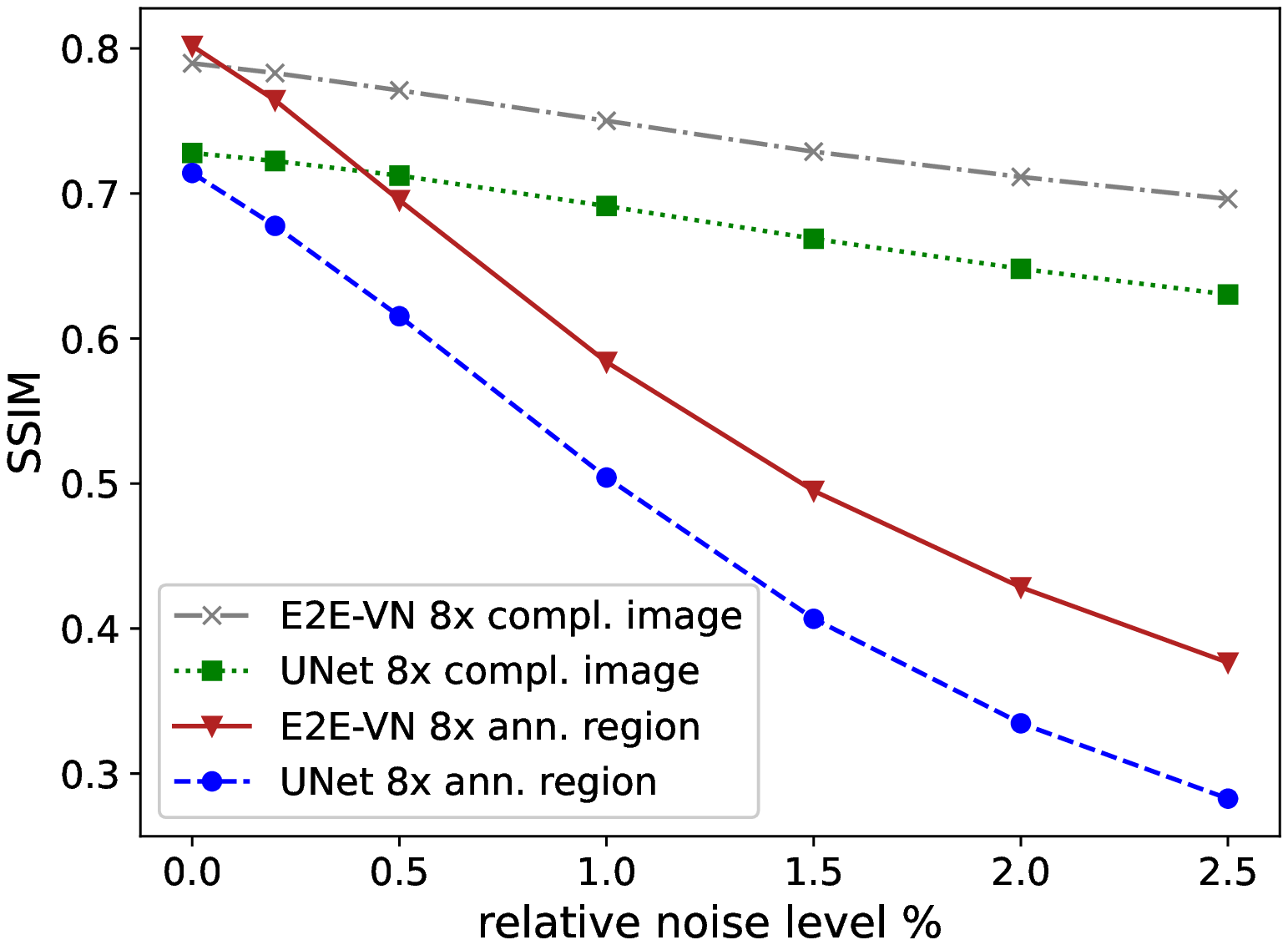}
         \caption{Annotated regions vs. entire image}
         \label{fig:PSNR-rel-error}
     \end{subfigure}
    \caption{SSIM between the full $k$-space reconstruction and DL-based reconstruction with adversarial noise. (a) shows the behaviour with respect to the acceleration factor, while (b) compares attacks on annotated, diagnostically relevant regions with attacks on the entire image.}
    \label{fig:noise_attack_plots}
\end{figure}

\begin{figure}[ht!]
     \centering
     \begin{subfigure}[b]{0.48\textwidth}
         \centering
         \includegraphics[width=\textwidth]{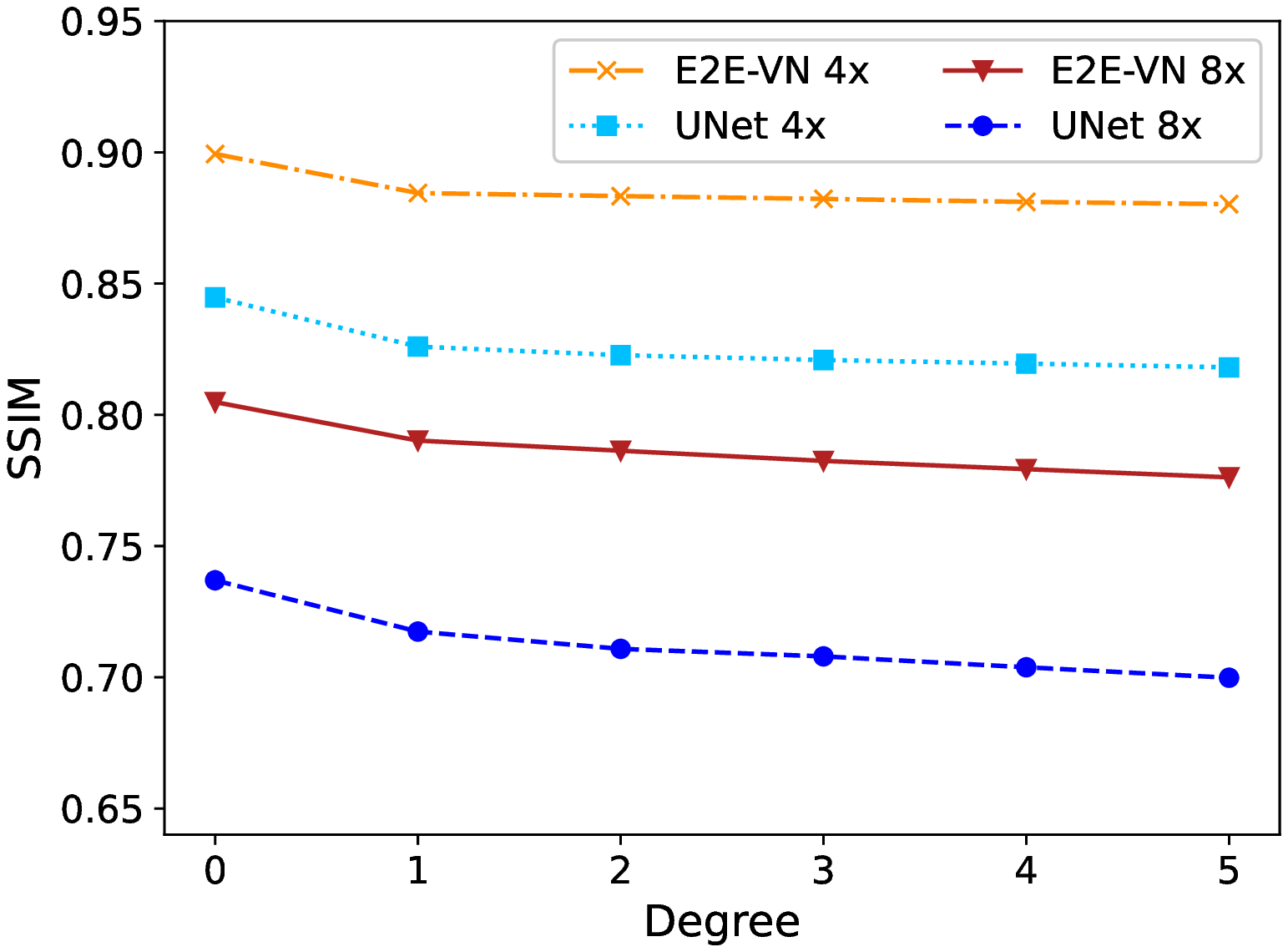}
         \caption{Robustness to rotations}
         \label{fig:SSIM-rot-error}
     \end{subfigure}
     \hfill
     \begin{subfigure}[b]{0.48\textwidth}
         \centering
         \includegraphics[width=\textwidth]{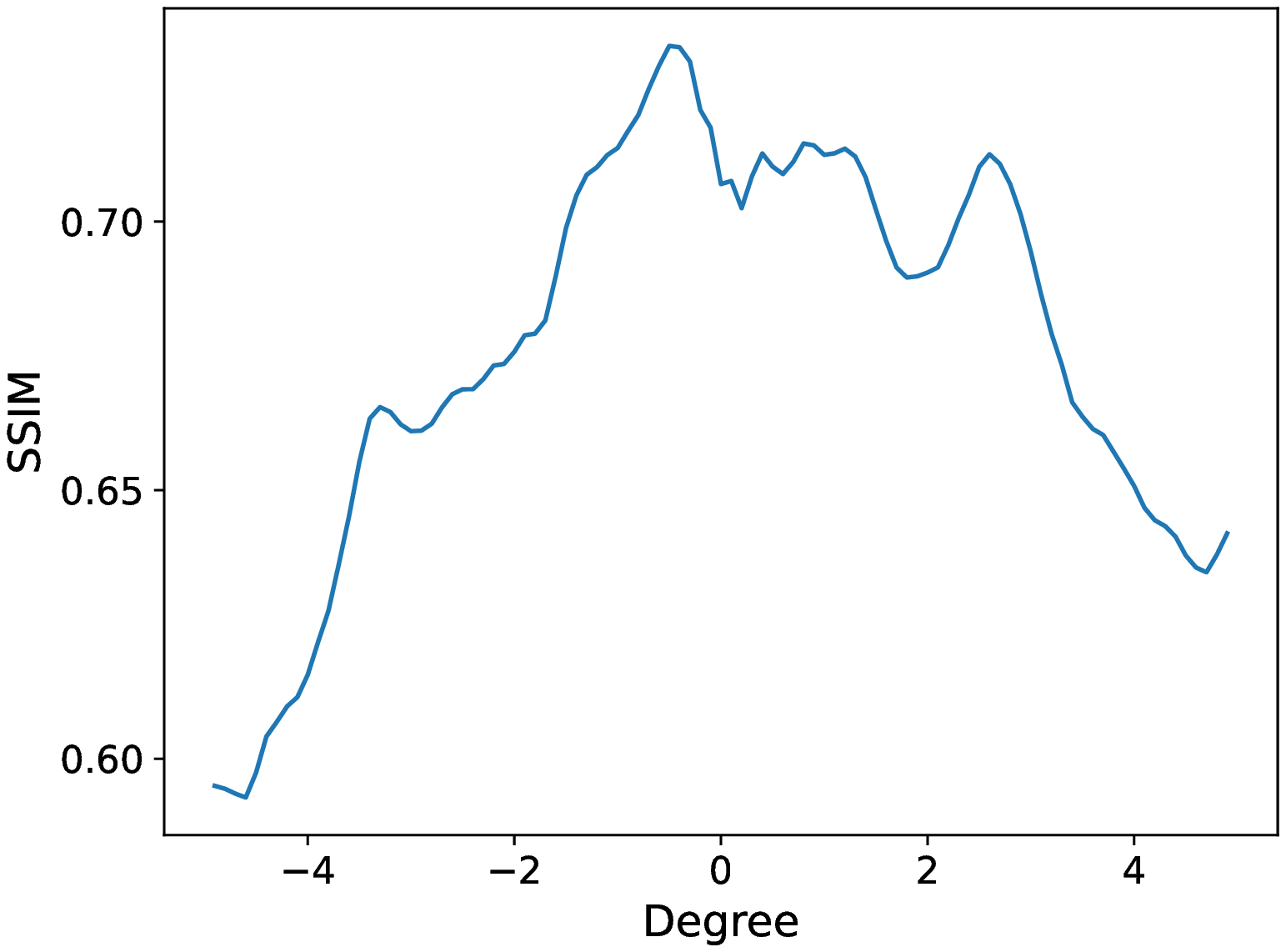}
         \caption{Worst case selection} %
         \label{fig:PSNR-rot-error}
     \end{subfigure}
    \caption{Results of adversarial rotation experiments. (a) contains the methods' SSIM with increasing severity of the attack. (b) visualises the sensitivity of the SSIM score with respect to $\theta$ for the qualitative example shown in Fig.~\ref{fig:rotation_effects}.}
    \label{fig:rotation_errors}
\end{figure}

Fig.~\ref{fig:noise_attack_plots} shows the relative degradation of the structural similarity (SSIM) computed over the diagnostically relevant regions for the noise attack model. Overall E2E-VN yields consistently better results compared to the simpler UNet-based reconstruction method. Nevertheless, the slope of the degradation is similar for both techniques, suggesting that both methods are equally susceptible to the attack. These findings also hold for the rotation attacks shown in Fig.~\ref{fig:SSIM-rot-error}. The similar sensitivity to adversarial attacks is surprising given that the E2E-VN is implicitly an iterative approach with enforced data consistency. 

\begin{figure}[ht!]
    \centering
    \includegraphics[width=0.75\linewidth]{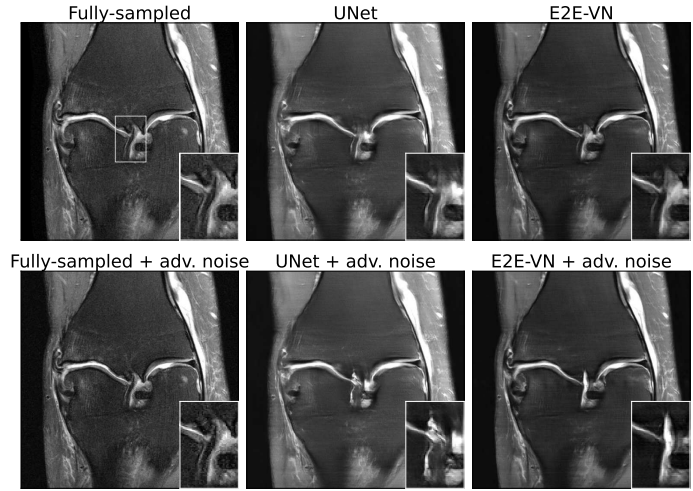}
    \caption{
     Comparison of reconstruction algorithms with a relative noise-level of 2\% and 8x-acceleration (top row: unperturbed reconstruction, bottom row: reconstruction with adversarial noise).
     The fully sampled image shows a low grade sprain of the anterior cruciate ligament (ACL). Both the UNet and the E2E-VN reconstructions with adv. noise could lead to clinically relevant misdiagnoses. Note that changes are barely visible if the adversarial noise is added to the fully-sampled $k$-space (bottom left).
    }
    \label{fig:reconstruction_w_noise_example}
\end{figure}

We further analysed the results qualitatively in collaboration with a senior radiologist. In Fig. \ref{fig:reconstruction_w_noise_example}, reconstructions of the UNet, the E2E-VN as well as reconstructions of both methods with adversarial noise are compared for a sample image. While the fully sampled reconstruction shows a low grade sprain of the anterior cruciate ligament (ACL), the UNet reconstruction with adversarial noise shows a thickening and distention, which could mistakenly be interpreted as an acute injury. The E2E-VN reconstruction with adversarial noise on the other hand could lead to a false diagnosis of an old ACL tear. It can thus be concluded that even small amounts of adversarial noise can lead to a false diagnosis with potential clinical consequences.

\begin{figure}[ht!]
    \centering
    \includegraphics[width=0.85\linewidth]{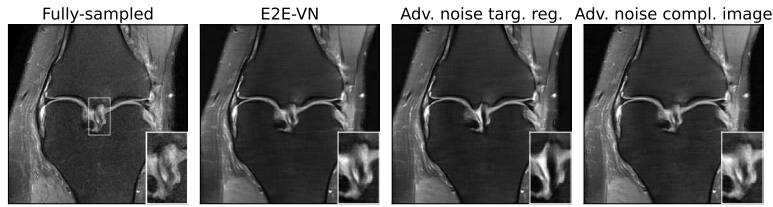}
    \caption{
    Comparison of attacking the targeted region only (third column) versus attacking the complete image (fourth column) using the E2E-VN trained for $8\times$ acceleration. Attacking only the annotated region, the ligament appears more homogeneous, which could lead to a misdiagnosis of the ACL rupture.
    }
    \label{fig:comparison_fullimage_targeted_region}
\end{figure}

Fig.~\ref{fig:PSNR-rel-error}
compares attacks targeted on annotated regions with attacks on entire images (the latter being the standard approach e.g. \cite{Antun,solving_inverse_problems,darestani2021measuring}). While \cite{solving_inverse_problems} reported that diagnostically relevant regions remain relatively unaffected by adversarial noise attacks, we are able to show that when the attack is targeting those regions directly, the effect is much more severe. A qualitative example demonstrating this effect is shown in Fig.~\ref{fig:comparison_fullimage_targeted_region}. While the annotated region noticeably changed when the adversarial attack was focused on it (third column), the effect is much smaller when the attack was performed on the complete image (fourth column). Clinically interpreting the results, we note that when only the annotated region is attacked (third column), it starts to look more defined and homogeneous. This could lead to a failure to detect the ACL rupture which is present in the image.

\begin{figure}[ht!]
    \centering
    \includegraphics[width=0.80\linewidth]{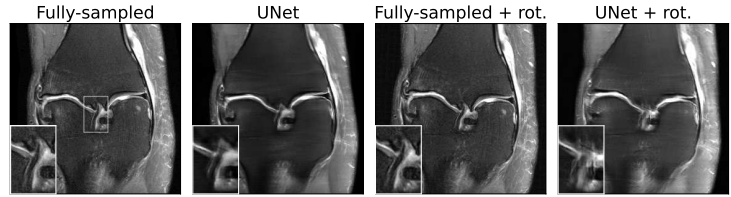}
    \caption{Effects of small rotation on the reconstruction quality (8$\times$ acceleration). For the worst-case angle of -4.9\degree, the annotated region gets more fuzzy, potentially leading to a misdiagnosis of an ACL tear instead of a low grade sprain. } 
    \label{fig:rotation_effects}
\end{figure}

The SSIM scores after adversarially attacking our models with rotations are shown in Fig.~\ref{fig:rotation_errors}. Comparing Fig.~\ref{fig:SSIM-rot-error} with Fig.~\ref{fig:SSIM-rel-error} above, it can be seen that rotation attacks are less severe than attacks with adversarial noise which is not unexpected given the much higher degree of freedom of the noise attacks. However, the rotation attacks, although very small and purportedly benign in nature, can still manage to substantially degrade the output of the algorithm. From Fig.~\ref{fig:PSNR-rot-error} it can be observed that the attack is continuous in $\theta$ and a non-negligible range of rotation angles leads to substantial drops in SSIM. This indicates that such degradation could realistically occur in clinical practice.  A qualitative example of a UNet reconstruction is shown in Fig.~\ref{fig:rotation_effects}. While the rotation has negligible effects on the fully-sampled reconstruction (third column), it can be seen that the UNet attacked with the same rotation (fourth column) produces an image where the ACL appears fuzzy (similar to the example in Fig.~\ref{fig:comparison_fullimage_targeted_region}). As in the previous examples, this could potentially lead to a misdiagnosis. 


\section{Conclusion}
In this work we have shown potential pitfalls of using DL-based MR-re\-con\-struc\-tion methods. By creating realistic adversarial changes to the $k$-space data we have shown that the reconstruction quality of common reconstruction methods declines significantly for both an iterative network (i.e. the E2E-VN) and a UNet-based approach. We observed that both methods, although the architecture and reconstruction method are quite different, encounter similar levels of reconstruction quality decline when adversarial perturbations are applied. By further targeting only specific regions we have shown that diagnostically relevant information can -- contrary to the findings of \cite{solving_inverse_problems} -- fall victim to adversarial attacks for the same noise levels. Lastly, we observed that both examined reconstruction algorithms are also sensitive to very small adversarial rotations in some instances also leading to loss of diagnostic information. We believe future work should increasingly incorporate such worst case analyses and focus on the creation of models that are relatively stable under such perturbations. A limitation of our work is that we explored reduced versions of the UNet and E2E-VN reconstruction architectures due to computational constraints. In future work, we aim to verify that our insights carry-over to the full size models. 


\section*{Acknowledgements}

Funded by the Deutsche Forschungsgemeinschaft (DFG, German Research Foundation)
under Germany’s Excellence Strategy – EXC number 2064/1 – Project number 390727645.
The authors thank the International Max Planck Research School for Intelligent Systems
(IMPRS-IS) for supporting Jan Nikolas Morshuis.

\bibliographystyle{splncs04}
\bibliography{paper18}

\begin{thebibliography}{10}
\providecommand{\url}[1]{\texttt{#1}}
\providecommand{\urlprefix}{URL }
\providecommand{\doi}[1]{https://doi.org/#1}

\bibitem{Antun}
Antun, V., Renna, F., Poon, C., Adcock, B., Hansen, A.C.: On instabilities of
  deep learning in image reconstruction and the potential costs of {AI}.
  Proceedings of the National Academy of Sciences  \textbf{117}(48),
  30088--30095 (2020). \doi{10.1073/pnas.1907377117},
  \url{https://www.pnas.org/doi/abs/10.1073/pnas.1907377117}

\bibitem{caliva2020adversarial}
Caliv{\'a}, F., Cheng, K., Shah, R., Pedoia, V.: Adversarial robust training of
  deep learning {MRI} reconstruction models. arXiv preprint arXiv:2011.00070
  (2020)

\bibitem{cheng2020}
Cheng, K., Caliv\'a, F., Shah, R., Han, M., Majumdar, S., Pedoia, V.:
  Addressing the false negative problem of deep learning {MRI} reconstruction
  models by adversarial attacks and robust training. In: Arbel, T., Ben~Ayed,
  I., de~Bruijne, M., Descoteaux, M., Lombaert, H., Pal, C. (eds.) Proceedings
  of the Third Conference on Medical Imaging with Deep Learning. Proceedings of
  Machine Learning Research, vol.~121, pp. 121--135. PMLR (06--08 Jul 2020),
  \url{https://proceedings.mlr.press/v121/cheng20a.html}

\bibitem{mraugment}
Fabian, Z., Heckel, R., Soltanolkotabi, M.: Data augmentation for deep learning
  based accelerated {MRI} reconstruction with limited data. In: Meila, M.,
  Zhang, T. (eds.) Proceedings of the 38th International Conference on Machine
  Learning. Proceedings of Machine Learning Research, vol.~139, pp. 3057--3067.
  PMLR (18--24 Jul 2021),
  \url{https://proceedings.mlr.press/v139/fabian21a.html}

\bibitem{firbank1999comparison}
Firbank, M., Coulthard, A., Harrison, R., Williams, E.: {A comparison of two
  methods for measuring the signal to noise ratio on MR images}. Physics in
  Medicine \& Biology  \textbf{44}(12), ~N261 (1999)

\bibitem{solving_inverse_problems}
Genzel, M., Macdonald, J., März, M.: Solving inverse problems with deep neural
  networks -- robustness included? (2020). \doi{10.48550/ARXIV.2011.04268},
  \url{https://arxiv.org/abs/2011.04268}

\bibitem{unet_reconstruction}
Hyun, C.M., Kim, H.P., Lee, S.M., Lee, S., Seo, J.K.: Deep learning for
  undersampled {MRI} reconstruction. Physics in Medicine {\&} Biology
  \textbf{63}(13),  135007 (jun 2018). \doi{10.1088/1361-6560/aac71a},
  \url{https://doi.org/10.1088/1361-6560/aac71a}

\bibitem{robustness_fastmri_2019}
Johnson, P.M., Jeong, G., Hammernik, K., Schlemper, J., Qin, C., Duan, J.,
  Rueckert, D., Lee, J., Pezzotti, N., De~Weerdt, E., Yousefi, S., Elmahdy,
  M.S., Van~Gemert, J.H.F., Sch{\"u}lke, C., Doneva, M., Nielsen, T.,
  Kastryulin, S., Lelieveldt, B.P.F., Van~Osch, M.J.P., Staring, M., Chen,
  E.Z., Wang, P., Chen, X., Chen, T., Patel, V.M., Sun, S., Shin, H., Jun, Y.,
  Eo, T., Kim, S., Kim, T., Hwang, D., Putzky, P., Karkalousos, D., Teuwen, J.,
  Miriakov, N., Bakker, B., Caan, M., Welling, M., Muckley, M.J., Knoll, F.:
  {Evaluation of the Robustness of Learned MR Image Reconstruction to
  Systematic Deviations Between Training and Test Data for the Models from the
  {fastMRI} Challenge}. In: Haq, N., Johnson, P., Maier, A., W{\"u}rfl, T.,
  Yoo, J. (eds.) Machine Learning for Medical Image Reconstruction. pp. 25--34.
  Springer International Publishing, Cham (2021)

\bibitem{liu2022medical}
Liu, X., Glocker, B., McCradden, M.M., Ghassemi, M., Denniston, A.K.,
  Oakden-Rayner, L.: {The medical algorithmic audit}. The Lancet Digital Health
   (2022)

\bibitem{noise_in_mri}
Macovski, A.: {{N}oise in {M}{R}{I}}. Magn Reson Med  \textbf{36}(3),  494--497
  (Sep 1996)

\bibitem{results_fastmri_2020}
Muckley, M.J., Riemenschneider, B., Radmanesh, A., Kim, S., Jeong, G., Ko, J.,
  Jun, Y., Shin, H., Hwang, D., Mostapha, M., Arberet, S., Nickel, D., Ramzi,
  Z., Ciuciu, P., Starck, J.L., Teuwen, J., Karkalousos, D., Zhang, C., Sriram,
  A., Huang, Z., Yakubova, N., Lui, Y.W., Knoll, F.: Results of the 2020
  {fastMRI} challenge for machine learning {MR} image reconstruction. IEEE
  Transactions on Medical Imaging  \textbf{40}(9),  2306--2317 (2021).
  \doi{10.1109/TMI.2021.3075856}

\bibitem{rss}
Roemer, P.B., Edelstein, W.A., Hayes, C.E., Souza, S.P., Mueller, O.M.: The
  {NMR} phased array. Magn Reson Med  \textbf{16}(2),  192--225 (Nov 1990)

\bibitem{unet_orig}
Ronneberger, O., Fischer, P., Brox, T.: {U-Net: Convolutional Networks for
  Biomedical Image Segmentation}. In: Navab, N., Hornegger, J., Wells, W.M.,
  Frangi, A.F. (eds.) Medical Image Computing and Computer-Assisted
  Intervention -- MICCAI 2015. pp. 234--241. Springer International Publishing,
  Cham (2015)

\bibitem{rudin1992nonlinear}
Rudin, L.I., Osher, S., Fatemi, E.: {Nonlinear total variation based noise
  removal algorithms}. Physica D: nonlinear phenomena  \textbf{60}(1-4),
  259--268 (1992)

\bibitem{schlemper2017deep}
Schlemper, J., Caballero, J., Hajnal, J.V., Price, A., Rueckert, D.: A deep
  cascade of convolutional neural networks for {MR} image reconstruction. In:
  International conference on information processing in medical imaging. pp.
  647--658. Springer (2017)

\bibitem{end-to-end-varnet}
Sriram, A., Zbontar, J., Murrell, T., Defazio, A., Zitnick, C.L., Yakubova, N.,
  Knoll, F., Johnson, P.: End-to-end variational networks for accelerated {MRI}
  reconstruction. In: Martel, A.L., Abolmaesumi, P., Stoyanov, D., Mateus, D.,
  Zuluaga, M.A., Zhou, S.K., Racoceanu, D., Joskowicz, L. (eds.) Medical Image
  Computing and Computer Assisted Intervention -- MICCAI 2020. pp. 64--73.
  Springer International Publishing, Cham (2020)

\bibitem{tezcan2018mr}
Tezcan, K.C., Baumgartner, C.F., Luechinger, R., Pruessmann, K.P., Konukoglu,
  E.: {MR} image reconstruction using deep density priors. IEEE transactions on
  medical imaging  \textbf{38}(7),  1633--1642 (2018)

\bibitem{darestani2021measuring}
Zalbagi~Darestani, M., S.~Chaudhari, A., Heckel, R.: Measuring robustness in
  deep learning based compressive sensing. In: International Conference on
  Machine Learning (ICML) (2021)

\bibitem{fastmri}
Zbontar, J., Knoll, F., Sriram, A., Murrell, T., Huang, Z., Muckley, M.J.,
  Defazio, A., Stern, R., Johnson, P., Bruno, M., Parente, M., Geras, K.J.,
  Katsnelson, J., Chandarana, H., Zhang, Z., Drozdzal, M., Romero, A., Rabbat,
  M., Vincent, P., Yakubova, N., Pinkerton, J., Wang, D., Owens, E., Zitnick,
  C.L., Recht, M.P., Sodickson, D.K., Lui, Y.W.: {fastMRI}: An open dataset and
  benchmarks for accelerated {MRI} (2018)

\bibitem{fastmri_plus}
Zhao, R., Yaman, B., Zhang, Y., Stewart, R., Dixon, A., Knoll, F., Huang, Z.,
  Lui, Y.W., Hansen, M.S., Lungren, M.P.: {fastMRI+}, clinical pathology
  annotations for knee and brain fully sampled magnetic resonance imaging data.
  Scientific Data  \textbf{9}(1), ~152 (Apr 2022).
  \doi{10.1038/s41597-022-01255-z},
  \url{https://doi.org/10.1038/s41597-022-01255-z}

\bibitem{zhu2018image}
Zhu, B., Liu, J.Z., Cauley, S.F., Rosen, B.R., Rosen, M.S.: Image
  reconstruction by domain-transform manifold learning. Nature
  \textbf{555}(7697),  487--492 (2018)

\end{thebibliography}

\end{document}